\documentclass[%
 reprint, showkeys,
nofootinbib,
 amsmath,amssymb,
 aps,
nolongbibliography
]{revtex4-2}

\usepackage{dcolumn}
\usepackage{bm}
\usepackage{physics}

\usepackage{physics}
\usepackage{mathtools}
\usepackage{amsmath, amssymb, amsthm, amsbsy, bbm, braket}
\usepackage{dsfont}
\usepackage[percent]{overpic}
\usepackage{xcolor}
\usepackage{enumitem}
\usepackage{slashed}

\usepackage[colorlinks]{hyperref}

\hypersetup{pdfstartview={XYZ null null 1.25},colorlinks=true,citecolor=blue,linkcolor=blue,urlcolor=blue,}

\def\l{\lambda}

\def\m{\mu}

\begin{document}

\title{Soft Scalars don't decouple} 

\author{Shovon Biswas}
 \email{shovon@phas.ubc.ca}
\author{Gordon W. Semenoff}%
 \email{gordonws@phas.ubc.ca}
\affiliation{%
Department of Physics and Astronomy University of British Columbia,
6224 Agricultural Road, Vancouver, British Columbia V6T 1Z1, Canada.
}


\begin{abstract}
It is demonstrated that it is possible to find a field theory containing massless scalar particles which has infrared structure closely resembling that of quantum electrodynamics and perturbative quantum gravity but exhibiting no gauge invariance or internal symmetries at all, and in particular, no apparent asymptotic symmetry. 
 It is shown that, unlike 
 soft  photons and gravitons, the soft scalars do not decouple from dressed states and they are generically produced when hard dressed particles interact. However,  the entanglement  of the hard and resulting soft particles is vanishingly small.
\end{abstract}

\maketitle


\section{Introduction}\label{sec:intro}

Infrared divergences \cite{Bloch1937, yennie1961,weinberg1965} due to a
massless real scalar field in four spacetime dimensions can be, for the most part, structurally identical  to those of the massless photons of quantum electrodynamics or the massless gravitons in the  perturbative effective field theory of quantum gravity near a flat background.   However, there is a big difference between scalar fields and photons or gravitons in that massless scalar fields can in principle occur in quantum field theories which do not exhibit any gauge invariance or even continuous or discrete global symmetry and therefore do not exhibit the asymptotic symmetries which seem intimately tied to the infrared problem of photons and gravitons \cite{Strominger}. In this paper, we shall study the infrared divergences and their cure in such a scalar field theory  and attempt to answer the question as to what is different in that case.  

What we will find, in the context of a very specific example, where there are no obvious internal symmetries at all,  is one big difference.  In this model, one can construct a dressed state  of a hard particle where the hard particle is accompanied by a cloud of soft scalars and the soft scalar content of the cloud is fine-tuned in such a way that scattering amplitudes for these dressed particles are free of infrared divergences. Of course such a dressing by soft photons or soft gravitons is already well known, and is already proposed as a solution of the infrared problem of quantum electrodynamics \cite{Chung1965,Kibble1968,Kulish1971} and  perturbative quantum gravity \cite{Ware2013}.  In those cases, once electrically and gravitationally charged particles are dressed, the dressing can be further fine-tuned in such a way that the soft photons and soft gravitons decouple completely.  The $S$ matrix factorizes into a hard sector and a soft sector.  Corrections to this factorization are suppressed by powers of an infrared cutoff, $E_{res}$. 

For the soft scalar fields in our model, this is not the case. There is a residual coupling of the scalar-dressed hard particles to soft scalars which is generically as important as the coupling of the dressed hard particles with each other.  There is no way to adjust the dressing in order to remove this interaction.  This means that even once the hard particles are dressed with soft scalars, an interaction of the hard particles will still produce more soft scalars.  

We find that  what does decouple is information.  Like the case of photons and gravitons, the interaction of undressed hard particles  produces infinite numbers of soft scalars.  
Moreover, the $S$ matrix for scattering processes involving hard particles is infrared divergent. We will confirm these facts  in the context of our scalar model. We will also  confirm that, like photons and gravitons, the soft scalars which escape detection in a scattering experiment carry away copious amounts of information to the point that their entanglement entropy with  the hard particles left behind is itself infrared divergent.   The result  of this entanglement, like for photons and gravitons, is decoherence of the final state of the hard particles and suppression of interference phenomena \cite{Carney2017,Carney2018,Carney2018a}.  

When the hard particles are dressed with soft scalars, on the other hand, even if they don't decouple and their interaction still produces some soft scalars, the entanglement between those scalars and the dressed hard particles is negligible, suppressed by powers of the infrared cutoff $E_{res}$.  Even when the soft scalars fly away from a scattering event undetected, the loss of quantum information, which would be manifest in decoherence, is negligible. In this sense it is information that decouples.  

 The rest of this paper is  an exposition of the results described in the paragraphs above.  It is organized as follows.  In section \ref{sec: soft_theorem}, we will discuss the details of the scalar field theory model that we will use and we will give  a derivation of the soft scalar theorem. The soft scalar theorem will  be used to study scattering amplitudes in later sections.   In  section \ref{canc} we will confirm that, for undressed states, the structure of infrared divergences is practically identical to those of photons or gravitons.  We will discuss how the infrared problem is addressed using the Bloch-Nordseick scheme of using inclusive probabilities and we will demonstrate that infrared divergences indeed cancel for the questions that are traditionally asked by particle physicists.  Then we will show that some other questions about certain interference phenomena or decoherence are still severely affected by the infrared divergences. 
  
  In section \ref{sec: dressing},  we will construct dressed states of hard particles.  There, we insist that the dressing is added by a formally unitary transformation.  This is important since, in the infrared cutoff theory, this is a unitary transform, a simple change of basis in the Hilbert space.  When the fundamental infrared cutoff is removed, it becomes an improper unitary transformation. However, even in that case, if it is formally unitary, it implements a canonical transformation between inequivalent representations of the operator algebra of the quantum field theory.   We also discuss the Faddeev-Kulish modification of the $S$ matrix that is to be computed to find the scattering amplitudes for dressed particles.   In section \ref{cancellation},  we demonstrate that the infrared singularities indeed cancel  from those amplitudes.
  In section \ref{sec: non-decoupling}, we establish that  soft scalars do not decouple from the dressed states. In section \ref{conclusion}, we discuss the implications.  We also argue that, even though soft scalars do not decouple, they  do not drive appreciable decoherence. 
\section{Soft Scalar Theorem} \label{sec: soft_theorem}

Consider a Majorana fermion of mass $m$ coupled to a massless   real scalar field $\phi$ via a Yukawa coupling in four spacetime dimensions. The Lagrangian density is given by
\begin{equation}\label{lagrangian_density}
    \mathcal{L}=-\frac{i}{2}\bar{\psi}\left[\slashed{\partial}+m-g\phi\right]\psi-\frac{1}{2}\partial_\mu \phi\;\partial ^\mu \phi - V(\phi)
\end{equation}
where $\psi$ is the Majorana spinor field, $g$ is the Yukawa coupling constant and $\phi$ is the scalar field. Since the Yukawa coupling is marginal in four dimensions, $g$ is dimensionless.  We will assume that we can always choose counter-terms so that the scalar field tadpole $\sim\phi$, the scalar field mass term $\sim \phi^2$ and the scalar field trivalent coupling $\sim \phi^3$ are all canceled exactly at each order of perturbation theory.   To make the Lehmann-Symanzik-Zimmermann (LSZ) reduction formulae simpler, we will also assume that the subtraction scheme can be chosen so that the pole in the scalar field propagator has unit residue, $\sim \frac{-i}{q^2-i\epsilon}+\ldots$.  
We note that this quantum field theory has no obvious internal symmetries at all. Fermion parity $(-1)^F$ can be regarded as a space time symmetry since it is a rotation by $2\pi$. Our use of the Majorana fermion is not essential as results would be very similar for a complex fermion, which would of course have at least one internal $U(1)$ symmetry. Our use of Majorana fermions is motivated by wanting a  model with no internal symmetry at all.  We will study  the  infrared divergences which occur in the analysis of scattering experiments involving the asymptotic particles of this quantum field theory. We will assume that the coupling is weak so that the particle spectrum resembles the tree level one, containing one Majorana fermion with mass $m$ and one massless real scalar field. There has already been some discussion of soft scalar theorems \cite{Cheung2022}, infrared divergences and 
the possibility of asymptotic symmetries playing a role for scalar fields \cite{Henneaux2019,Campiglia2018,Campiglia2018a}.  The latter has been discussed in the context of a dual antisymmetric tensor gauge field representation which, for a Yukawa coupling like we use here, is not related to this theory by a local transformation.   This is not the direction  that we will pursue here. Instead we will examine the behaviour of the scattering matrix in the quantized theory  corresponding to (\ref{lagrangian_density}).
 
 A scattering experiment has an incoming state.  We will use the notation $|\{p\}\{k\}\{q\}\rangle$ for such a state, with fermions having momenta and helicity in a set $\{p\}$, hard scalars with momenta  $\{k\}$ and soft scalars with momenta $\{q\}$.  In the course of the scattering, an incoming state  evolves to  an outgoing state which is a quantum superposition of the incoming states, 
\begin{align}
& |\{p\}\{k\}\{q\}\rangle\implies
\nonumber \\ 
 &\sum_{\{p'\}\{k'\}\{q'\}} |\{p'\}\{k'\}\{q'\}\rangle~S^\dagger( \{p'\}\{k'\}\{q'\} ;  \{p\}\{k\}\{q\} )
\label{scattering}
\end{align}
The coefficients in the superposition, $~S^\dagger( \{p'\}\{k'\}\{q'\} ;  \{p\}\{k\}\{q\} )$,  are elements of the $S$ matrix. The dagger in the above formula is there to match conventions,
\begin{align}\label{S}
&S^\lambda( \{p\}\{k\}\{q\}; \{p'\}\{k'\}\{q'\} ) \nonumber\\
&=\langle\{p\}\{k\}\{q\}|
~S^\lambda~|\{p'\}\{k'\}\{q'\} \rangle
\end{align}
We have added a superscript $\lambda$  to the $S$ matrix  to remind ourselves that its definition requires a fundamental infrared cutoff.  
We will denote this fundamental cutoff by $\lambda$.\footnote{An easy, Lorentz invariant way of introducing such a cutoff is to simply allow the scalar field to have a small mass. }
    With the cutoff taken into account,  the  operator $S^\lambda $ whose matrix elements are discussed above  is the Dyson $S$ matrix which is computed using the usual LSZ reduction formulae, time dependent perturbation theory and Feynman diagrams.  In the rest of this paper, what we mean by  an infrared finite quantity is that said quantity remains finite as we take $\lambda\to 0$. The matrix elements of $S^\lambda$ in (\ref{S}) are generally not such a quantity.

Beyond the fundamental infrared cutoff $\lambda$, we shall also require a distinction between hard and soft particles.  Any particle which has energy  above a threshold $E_{res}$ will be called a hard particle.  Any particle which has energy less than $E_{res}$ will be called a soft particle.  We shall call $E_{res}$ the ``detector resolution'' .  We will require a hierarchy of scales 
\begin{align}\label{hierarchy}
\lambda\ll E_{res},\Lambda
\ll m,~{\rm energies~ of~ hard~ particles}
\end{align}
where $\Lambda$ is a third infrared cutoff, distinct from $\lambda$ and $E_{res}$ that we will introduce shortly. 
Fermions are always hard particles.  The massless scalar, on the other hand can either be hard or soft, depending on its energy.  The validity of the  arguments in the rest of this paper will need the inequalities in equation (\ref{hierarchy}).

Let us consider an amplitude for the scattering of some fermions and hard scalar particles as well as the production  of a soft scalar particle.  Whenever this occurs, there are some contributions to the amplitude which are singular as the momentum of the soft scalar approaches zero. 
 In direct analogy with the same phenomenon in quantum electrodynamics, which is outlined in beautiful detail in Weinberg's book \cite{Weinberg2005}, the most singular parts come from the emission of the soft scalar from external lines of the amplitude. This singular part,  $\sim\frac{1}{q}$ for soft scalar with momentum $q$,  
 and the next-to leading behaviour 
 due to a scalar emitted from an outgoing fermion line with momentum $p'$,  is gotten from the amplitude without the scalar emission by making the replacement

 \begin{align}\label{outgoing_soft}
 &\bar {u}_r(p') ~~\to~~   \lim_{q\rightarrow 0}\;\bar {u}_r(p')(-g)\frac{-1}{i\slashed{p}'+i\slashed{q}+m-i\epsilon}\nonumber\\
 &=  \lim_{q\rightarrow 0}\;\bar {u}_r(p')\bigg(\frac{2gm}{p'^2+2p'\cdot q+q^2+m^2-i\epsilon}\nonumber\\
 &~~~~~~~~~~~~~~~~~~~~~~~~~~~-\frac{g}{-i\slashed{p}'-i\slashed{q}+m-i\epsilon}\bigg)\nonumber\\
 & =\left(\frac{g m}{p'\cdot q-i\epsilon}-\frac{g}{2m}\right) \bar{u}_r(p')  +\mathcal O(q)
\end{align}
where $ \bar {u}_r(p)$ is the momentum space spinor wave-function of the outgoing fermion and 
we have used the fact that the spinor satisfies the 
Dirac equation $\bar{u}_r(p')(i\slashed{p}'+m)=0 $.  Here, we note that the contribution to the next-to leading behaviour  from the external fermion line has a very simple form for a scalar field. Corrections to this formula go to zero as $q$ goes to zero.

Similarly, when the soft scalar emission is from an incoming fermion line, the singular part and next-to-leading  contribution is  
\begin{align}\label{incoming_soft}
u_r(p)~\to~ u_r(p)~  \left(   \frac{g m}{-p\cdot q-i\epsilon} - \frac{g}{2m}\right)+\mathcal O(q)
\end{align}

Of course, soft emissions can also take place from internal lines in the Feynman diagrams which contribute to the amplitude. The contribution of such processes is not singular at small $q$, but it does compete with the next-to-leading terms in (\ref{outgoing_soft}) and (\ref{incoming_soft}).   The effect of one soft emission from an internal line is the same as adding a vertex, together with a vertex counter-term to each of the fermion propagators inside the amputated correlation function that is used to form the $S$ matrix element $S^\lambda(\{p\}\{k\}; \{p'\}\{k'\} ) $.  We will denote this process by the symbol $-g\hat\partial_m$. We note that  the operation $\hat\partial_m$ is related to but not exactly the same as taking the derivative of amputated correlation function by the renormalized fermion mass $m$.  The discrepancy is due to the fact that we need a subtraction scheme where some counter-terms are $m$-dependent in a way that does not preserve the $m-g\phi$ structure that appears in the Lagrangian (\ref{lagrangian_density}). It is easy to see that even though the $m-g\phi$ structure can be maintained at the tree level, it is already violated at the one-loop level.

Putting this together, and a similar one for soft scalar absorption, we have the leading and next-to-leading soft scalar theorem

\begin{align}
S^\lambda(&\{p\},\{k\}; \{p'\}\{k'\},q' ) \nonumber\\
&= \left\{\sum_{p_n\in\{p\}\{p'\} }\left(\frac{\eta_n g m  }{p_n\cdot q'-i\eta_n\epsilon}- \frac{  g}{2m}\right)- g\hat\partial_m\right\}\nonumber\\
&\times
    S^\lambda(\{p\}\{k\}; \{p'\}\{k'\} ) 
   +\mathcal O(q') 
  \label{sub_soft_theorem}
  \end{align}
  \begin{align}
  S^\lambda(&\{p\},\{k\},q; \{p'\}\{k'\})
  \nonumber\\
  &= \left\{\sum_{p_n\in\{p\}\{p'\} }\left(\frac{\eta_n g m  }{-p_n\cdot q-i\eta_n\epsilon}- \frac{  g}{2m}\right)- g\hat\partial_m\right\}\nonumber\\
  &\times S^\lambda(\{p\}\{k\}; \{p'\}\{k'\} )
   +\mathcal O(q) 
  \label{sub_soft_theorem_1} 
  \end{align}
where $\eta_n=+1(-1)$ if $p_n$ is the momentum of an outgoing (incoming) line. 
   Now let us consider the amplitude for a process where an incoming state of hard particles and $M$ additional soft scalars $|\{p\}\{k\},q_1,\ldots,q_M\rangle$   evolves to another  state of hard particles  but with $N$ additional soft particles $|\{p'\}\{k'\},q_1',\ldots,q_N'\rangle$.  Using the leading parts of the soft scalar theorem in equations (\ref{sub_soft_theorem}) and (\ref{sub_soft_theorem_1}), it follows that the most singular part of the $S$ matrix element  is  given by\footnote{We have not attached momentum space wave-functions for the soft scalars.  We will take the convention of including them as the appropriate factors in the scalar states. } 
 \begin{align}
 & S^\lambda(\{p\}\{k\},q_1,\ldots,q_M; \{p'\}\{k'\}, q_1',...q_N')\nonumber \\  &=  S^\lambda(\{p\}\{k\}; \{p'\}\{k'\} ) 
 \prod_{j=1}^M\left( \sum_{p_n\in\{p\}\{p'\}}\frac {\eta_n g m }{-p_n \cdot q_j-i\eta_n\epsilon}\right)\nonumber\\
& \times
 \prod_{r=1}^N\left( \sum_{p_n\in\{p\}\{p'\}}\frac {\eta_n g m }{p_n \cdot q_r'-i\eta_n\epsilon}\right)  +\ldots
    \label{soft1}
\end{align}
where the ellipses denote terms less singular at small $q$ or $q'$ than $\frac{1}{q^M}\frac{1}{{q'}^N}$. 
 The remarkable fact about this soft  theorem is that it gives us the most important part of the soft scalar production amplitude for any process if we know the amplitude of the process without the soft scalar production.  What is more, it is practically identical to the one for soft photons or gravitons where only the numerators in the singular factors ( here it is $gm$) are slightly different.


\section{Infrared divergence cancellation in the Bloch-Nordseick Scheme}\label{canc}

 It is easy to see that, in direct parallel with quantum electrodynamics and perturbative quantum gravity, the infrared divergences in the $S$ matrix itself come from loop integrals where both ends of a single scalar propagator in the loop end on  external fermion lines.  Introduction of the fundamental infrared cutoff $\lambda$ renders these loop integrals finite and of order $\sim \ln(\lambda)$. 
 
 We imagine that, in these loop integrals, the infrared cutoff $\lambda$ could be replaced by a more convenient one which we shall call $\Lambda$.  This is a third infrared cutoff, distinct from $E_{res}$ and $\lambda$. with the assumption that it is much larger than the fundamental cutoff, $\Lambda\gg\lambda$, but it would still work as an infrared cutoff in that $\Lambda\ll m$ and that $\Lambda$ is much smaller than the momentum scales of any of the hard particles. It is in the same interval of the hierarchy (\ref{hierarchy}) as the detector resolution $E_{res}$.   Then with the new cutoff the logarithmically divergent integral goes as $\sim \ln(\Lambda)$ and the original integral with cutoff $\lambda$ goes like $\sim\ln(\lambda)= \ln (\Lambda) + \ln (\lambda/\Lambda)$, the first logarithm being produced by the integration over loop momenta from $\Lambda$ to infinity and the second being produced by the integration over loop momenta between $\lambda$ and $\Lambda$.  
Moreover, it is well known how to separate and sum up the latter, $ \ln (\lambda/\Lambda)$ contributions.   We refer the reader to Weinberg's book \cite{Weinberg2005} for the details in the case of photons and we note that, for massless scalars, the argument is practically identical.  The result is a relationship between the $S$ matrix defined with the two different infrared cutoffs
\begin{widetext}
\begin{align}
S^\lambda( &\{p\}\{k\}; \{p'\}\{k'\} )
\nonumber\\
&=S^\Lambda( \{p\}\{k\}; \{p'\}\{k'\} )e^{-i \bar\Phi(\{p\})-i\bar \Phi(\{p'\})}  \exp\biggl\{- \frac{1}{2}\int_\lambda^\Lambda\frac{d^3q}{(2\pi)^32|\vec q|}
\sum_{p_n\in\{p\}\{p'\}}\frac{gm\eta_n}{p_n\cdot q}\sum_{p_m\in\{p\}\{p'\}}\frac{gm\eta_m}{p_m\cdot q} \biggr\}
   \label{infrared_cutoff_s_matrix} \\
&=S^\Lambda( \{p\}\{k\}; \{p'\}\{k'\} ) \left(\frac{\lambda}{\Lambda}\right)^{\frac{1}{2}\bar A(\{p\},\{p'\})}e^{-i\bar \Phi(\{p\})-i\bar \Phi(\{p'\})}\, ;
   \label{infrared_cutoff_s_matrix_1}\\
&\bar A(\{p\},\{p'\})=-\frac{1}{8\pi^2}\sum_{p_np_m\in\{p\}\{p'\}}\frac{g^2m^2\eta_n\eta_m}{\gamma_{nm}}\ln\left(\frac{1+\xi_{nm}}{1-\xi_{nm}}\right), ~~\bar \Phi(\{p\})=\frac{1}{8\pi}\sum_{\substack{p_mp_n\in\{p\}\\m\neq n}}\frac{g^2m^2}{\gamma_{nm}} ~\ln\frac{\Lambda}{\lambda}\,,\\ &\xi_{nm}=\sqrt{1-\frac{m^4}{(p_n \cdot p_m)^2}}~,~~
 \gamma_{nm}=(p_n \cdot p_m)\sqrt{1-\frac{m^4}{(p_n \cdot p_m)^2}}\;.
 \label{phase}
\end{align} 
\end{widetext} 

To be clear, the beautiful formula (\ref{infrared_cutoff_s_matrix}) does not remove the fundamental infrared cutoff.  It simply gives us a relationship between $S$ matrices computed with different infrared cutoffs.  Furthermore, it is strictly valid only when
$
\lambda \ll \Lambda \ll m 
$.
We note that the exponent, $\bar{A}(\{p\},\{p'\})$, of the ratio of cutoffs contains data about the incoming and outgoing fermions only.  
It is independent of the   hard incoming or outgoing scalar particles. The phases, $\bar{\Phi}(\{p\}), \bar{\Phi}(\{p'\})$ are separated into two functions, one of incoming and one of outgoing fermion momenta.  Notice that the two functions have the same sign. 
  
For reasons which will become clear shortly, it is useful to express the evolution from the initial to final state in equation (\ref{scattering}) 
in the language of density matrices where a more general incoming state composed entirely of hard particles would be 
\begin{align*}
\sum_{ \substack{ \{p\}\{k\}\\ \{\tilde p\}\{\tilde k\}  } }~~ |\{p\}\{k\}\rangle\rho_{ \{p\}\{k\}\{\tilde p\}\{\tilde k\} }\langle\{\tilde p\}\{\tilde k\}| 
\end{align*}
with $\rho_{\{p\}\{k\}\{\tilde p\}\{\tilde k\} }$ an incoming density matrix.  In the course of a scattering experiment, this
incoming density matrix evolves to an outgoing one where the evolution is governed by the $S$ matrix,  
\begin{widetext}
\begin{align}
\sum_{ \substack{ \{p\}\{k\}\\ \{\tilde p\}\{\tilde k\}  } }~ |\{p\}\{k\}\rangle\rho_{\{p\}\{k\};\{\tilde p\}\{\tilde k\}}\langle\{\tilde p\}\{\tilde k\}|&\implies
\sum_{ \substack{ \{p\}\{k\}\\ \{\tilde p\}\{\tilde k\}  } } \rho_{\{p\}\{k\}\{\tilde p\}\{\tilde k\}} \sum_{\substack{ \{p'\}\{k'\}\{q'\} \\ \{\tilde p'\} \{\tilde k'\}\{\tilde q'\} } }  |\{p'\}\{k'\}\{q'\}\rangle\langle\{\tilde p'\}\{\tilde k'\}\{\tilde q'\}| 
\nonumber \\ &
 ~~~~~~~~ \times S^{\lambda\dagger}(  \{p'\}\{ k'\}\{q'\} ; \{p\}\{k\})~~
  S^\lambda( \{\tilde p\}\{\tilde k\} ;   \{\tilde p'\}\{\tilde k'\}\{\tilde q'\} ) \,.
\label{scattering_1}
\end{align}
\end{widetext}
We have assumed that there are no soft particles in the incoming states. However, soft particles are produced when the hard particles interact and they must appear in the final state density matrix. 
 We remember that the $S$ matrix elements in the expression above are infrared divergent and they are defined with a fundamental infrared cutoff, $\lambda$. 
We  will implement the  Bloch-Nordsieck mechanism where we make the assumption that, due to the limitations of detector resolution, the soft particles which are produced by the scattering are unobservable. They  fly away from the scattering experiment undetected.  What is left behind are 
the hard particles.  All of the experimentally accessible properties of the quantum state of the hard particles which remain are embedded in the reduced density matrix that is gotten from the final state density matrix in (\ref{scattering_1}) by taking a trace over all of the soft scalar states. 
The  reduced density matrix of the final state is thus
\begin{widetext}
\begin{align}
&\rho_{\rm final}= \sum_{ \substack{ \{p\}\{k\}\\ \{\tilde p\}\{\tilde k\}  } }~\rho_{\{p\}\{k\};\{\tilde p\}\{\tilde k\}}~~ \sum_{\substack{ \{p'\}\{k'\}  \\ \{\tilde p'\} \{\tilde k'\}  } }  |\{p'\}\{k'\}\rangle\langle\{\tilde p'\}\{\tilde k'\}| 
 \sum_{\{q\}} S^{\lambda\dagger}(  \{p'\}\{ k'\}\{q\} ; \{p\}\{k\})
  S^\lambda( \{\tilde p\}\{\tilde k\} ;   \{\tilde p'\}\{\tilde k'\}\{ q\} ) 
\label{scattering_2}
\end{align}
where $\sum_{\{q\}} $ denotes integration and summation over all possible soft scalar states. 
We can use the soft scalar theorem (\ref{soft1})  to simplify equation (\ref{scattering_2}). To take the trace, we identify pairs of ingoing and outgoing $q$'s and we integrate each identified pair over all values of $\vec q$ with $\lambda<|\vec q|<E_{res}$.  Using the soft scalar theorem yields the expression\footnote{Since, once the scalar momenta $q$ are on-shell, $p_n\cdot q>0$ and we can drop the $i\epsilon$'s from the denominators.}
\begin{align}
& \sum_{\{q\}} S^{\lambda\dagger}(  \{p'\}\{ k'\}\{q\} ; \{p\}\{k\})~~
  S^\lambda( \{\tilde p\}\{\tilde k\} ;   \{\tilde p'\}\{\tilde k'\}\{ q\} ) \nonumber \\
 &= \sum_{N=0}^\infty\frac{1}{N!}\int_\l^{E_{res}}\frac{d^3\vec{q_1}}{(2\pi)^32|\vec{q_1}|}\ldots\frac{d^3\vec{q_N}}{(2\pi)^32 |\vec{q_N}|} 
   \prod_{r=1}^N\left(   \sum_{ p_n\in\{p\}\{p'\}}\frac{gm\eta_n}{p_n\cdot q}
     \sum_{ p_n\in\{p\}\{p'\}} \frac{gm\eta_m}{-p_m\cdot q}  \right)
 \nonumber \\ &~~~~~~~~~~~~~~~ \times S^{\lambda\dagger}(  \{p'\}\{ k'\} ; \{p\}\{k\})~~
  S^\lambda( \{\tilde p\}\{\tilde k\} ;   \{\tilde p'\}\{\tilde k'\} )\,. \nonumber
  \end{align}
  
\end{widetext} 

  Notice that, in each factor in the product of integrals in the equation above, the three dimensional integration over $\vec q$ is over a narrow shell with $\lambda<|\vec q|<E_{res}$ and the result would be small, $\sim E_{res}^2$, if it were not for the singular terms due to soft scalar emission.  It is those singular terms which allow the integrals to be appreciable, in fact logarithmically infrared divergent $\sim \ln E_{res}/\lambda$.  Corrections to the above formula due to the non-singular next-to-leading contributions to the soft scalar theorem would be relatively suppressed by positive powers of $E_{res}$. 
  The summation in the equation above exponentiates and we find the expression for the reduced final state density matrix
\begin{widetext}
  \begin{align}
\rho_{\rm final}&=  \sum_{ \substack{ \{p\}\{k\}\\ \{\tilde p\}\{\tilde k\}  } }~ \rho_{\{p\}\{k\};\{\tilde p\}\{\tilde k\}} \sum_{\substack{ \{p'\}\{k'\}  \\ \{\tilde p'\}\{\tilde k'\}  } }  |\{p'\}\{k'\}\rangle\langle\{\tilde p'\}\{\tilde k'\}|~   S^{\lambda\dagger}(  \{p'\}\{ k'\} ; \{p\}\{k\})
  S^\lambda( \{\tilde p\}\{\tilde k\} ;   \{\tilde p'\}\{\tilde k'\} )  
\nonumber \\ &~~
 \times  \exp\biggl\{ \int_\lambda^{E_{res}}\frac{d^3\vec{q}}{(2\pi)^32|\vec{q}|} 
 \sum_{  p_n\in\{p\}\{p'\}  } \frac{gm\eta_n}{p_n\cdot q} 
 \sum_{ p_m\in\{\tilde p\}\{\tilde p'\}  } 
  \frac{ g m \eta_{m}  }{-p_{m}\cdot q}
 \biggr\}\,.
  \label{sscattering_3} 
  \end{align}
  
   Now, we must examine the infrared cutoff dependence of the $S$ matrix elements on the right-hand-side of equation (\ref{sscattering_3}). For this, we must use the $\lambda$-dependence of the $S$ matrix elements that is summarized in equation (\ref{infrared_cutoff_s_matrix}).  We then get  
  \begin{align}
&\rho_{\rm final}= \sum_{ \substack{ \{p\}\{k\}\\ \{\tilde p\}\{\tilde k\}  } }~\rho_{\{p\}\{k\}\{\tilde p\}\{\tilde k\}}~ \sum_{\substack{ \{p'\}\{k'\}  \\ \{\tilde p'\} \{\tilde k'\}  } } |\{p'\}\{k'\}\rangle\langle\{\tilde p'\}\{\tilde k'\}|  \nonumber \\ & \times
S^{\Lambda\dagger}(  \{p'\}\{ k'\} ; \{p\}\{k\})
  S^\Lambda( \{\tilde p\}\{\tilde k\} ;   \{\tilde p'\}\{\tilde k'\} ) 
e^{i \bar\Phi(\{p\}) + i\bar \Phi(\{p'\}) - i \bar\Phi(\{\tilde p\}) - i\bar \Phi(\{\tilde p'\})} \nonumber \\ & 
 \times  \exp\biggl\{ \int_\lambda^{E_{res}}\frac{d^3\vec{q}}{(2\pi)^32|\vec{q}|} 
 \sum_{  p_n\in\{p\}\{p'\}  } \frac{gm\eta_n}{p_n\cdot q}  
 \sum_{ p_m\in\{\tilde p\}\{\tilde p'\}  } 
  \frac{ g m \eta_{m}  }{-p_{m}\cdot q}
 \biggr\}
 \nonumber \\ &
\times  \exp\biggl\{- \int_\lambda^\Lambda\frac{d^3q}{(2\pi)^32|\vec q|}\biggl[ \frac{1}{2}
\biggl[ \sum_{ p_n\in\{  p\}\{  p'\}}  \frac{gm\eta_n}{p_n\cdot q}    \biggr]^2 +\frac{1}{2}
 \bigg[ \sum_{ p_n\in\{\tilde p\}\{\tilde p'\}}  \frac{gm\eta_n}{p_n\cdot q} \biggr]^2
  \biggr] \biggr\}\,.
     \label{scattering_5} 
  \end{align}
Upon combining the exponentials in the last two lines, we can write the above expression as
  \begin{align}
&\rho_{\rm final}=\sum_{ \substack{ \{p\}\{k\}\\ \{\tilde p\}\{\tilde k\}  } }~\rho_{\{p\}\{k\}\{\tilde p\}\{\tilde k\}}~~ \sum_{\substack{ \{p'\}\{k'\}  \\ \{\tilde p'\} \{\tilde k'\}  } } |\{p'\}\{k'\}\rangle\langle\{\tilde p'\}\{\tilde k'\}| \nonumber \\ & \times
S^{\Lambda\dagger}(  \{p'\}\{ k'\} ; \{p\}\{k\})
  S^\Lambda( \{\tilde p\}\{\tilde k\} ;   \{\tilde p'\}\{\tilde k'\} )  
e^{i \bar\Phi(\{p\}) + i\bar \Phi(\{p'\}) - i \bar\Phi(\{\tilde p\}) - i\bar \Phi(\{\tilde p'\})} \nonumber \\ & 
 \times  \exp\biggl\{ \int_\Lambda^{E_{res}}\frac{d^3\vec{q}}{(2\pi)^32|\vec{q}|} 
 \sum_{  p_n\in\{p\}\{p'\}  } \frac{gm\eta_n}{p_n\cdot q}  
 \sum_{ p_m\in\{\tilde p\}\{\tilde p'\}  } 
  \frac{ g m \eta_{m}  }{-p_{m}\cdot q}
 \biggr\}
 \nonumber \\ &
\times  \exp\biggl\{-\frac{1}{2} \int_\lambda^\Lambda\frac{d^3q}{(2\pi)^32|\vec q|}\biggl[  \sum_{ p_n\in\{  p\}\{  p'\}}  \frac{gm\eta_n}{p_n\cdot q}- \sum_{p_m \in\{ \tilde p\}\{ \tilde p'\}}  \frac{gm\eta_m}{p_m\cdot q} \biggr]^2\biggr\}\,.
     \label{scattering_7} 
  \end{align}
\end{widetext}
 Now we want to examine the right-hand-side of equation (\ref{scattering_7}) as the fundamental infrared cutoff $\lambda\to 0$. 
 The second line contains phases which are separately infrared divergent and we reserve comment on them for later.
 The third line is $\lambda$-independent and  infrared finite.
 The fourth (last) line has a negative semi-definite exponent which  can be written as
 \begin{align*}
 &\exp\biggl\{-\frac{1}{2} \int_\lambda^\Lambda\frac{d^3q}{(2\pi)^32|\vec q|}\biggl[  \sum_{ p_n\in\{  p\}\{  p'\}}  \frac{gm\eta_n}{p_n\cdot q}\nonumber\\
 &~~~~~~~~~- \sum_{p_m \in\{ \tilde p\}\{ \tilde p'\}}  \frac{gm\eta_m}{p_m\cdot q} \biggr]^2\biggr\}
 \\ &
=\exp\biggl\{-\frac{1}{32\pi^3} \ln\frac{\Lambda}{\lambda}\cdot\int  d\hat q \biggl[  \sum_{ p_n\in\{  p\}\{  p'\}}  \frac{gm\eta_n}{p_n\cdot v}\nonumber\\&~~~~~~~~- \sum_{p_m \in\{ \tilde p\}\{ \tilde p'\}}  \frac{gm\eta_m}{p_m\cdot v} \biggr]^2\biggr\}
\;;\; v^\mu=(1,\hat q)~,~~\hat q\equiv\vec q/|\vec q|\,.
  \end{align*}
 This exponent is either negative and logarithmically divergent or it vanishes. It can vanish only if the integrand in the integration over unit vectors vanishes, that is, if
\begin{align}
 \sum_{ p_n\in \{p\} \{\tilde p'\} }\frac{gm}{p_n\cdot v} =
\sum_{ p_m\in \{p'\} \{\tilde p\}  }\frac{gm}{p_m\cdot v}\,.
\label{condition}
\end{align}
It is only in this case where the $|\{p'\}\{k'\}\rangle\langle\{\tilde p'\}\{\tilde q'\}|$ matrix element of the reduced outgoing density matrix can be nonzero when the matrix 
element  $|\{p\}\{k\}\rangle\langle\{\tilde p\}\{\tilde q\}|$ of the incoming density matrix was nonzero.

   Remember that the    sums in (\ref{condition})  are over terms containing hard fermion momenta only.  Hard scalar momenta do not enter in these expressions. 
 The above equation must be so for all values of the null four-vector $v^\mu = \left(1,\hat q\right)$. If we Taylor expand the  above in powers off $\frac{\vec p_n}{\sqrt{\vec p_n^2+m^2}}<1$ and equate each order we see that 
 \begin{align*}
 \sum_{p_n\in \{p\}\{\tilde p'\}} \frac{(\hat q\cdot \vec p_n)^\ell }{(\sqrt{\vec p_n^2+m^2})^{\ell+1}}
&=
 \sum_{p_n\in \{\tilde p\}\{p'\} } \frac{(\hat q \cdot \vec p_n)^\ell }{(\sqrt{\vec p_n^2+m^2})^{\ell+1}},\nonumber\\
 &~~~~~~~~~~~~~~~~~~~~~~~~\forall\ell,~\forall \hat q\,. \end{align*}
 which implies that all multipole moments of the set of fermion momentum vectors $\{p\}\cup \{\tilde p'\}$ are equal to all moments of  the set $\{p'\}\cup \{\tilde p\}$ which can only be so if
 the two sets of vectors are identical 
 \begin{align}
 \{p\}\cup \{\tilde p'\}=\{p'\}\cup \{\tilde p\}\,.
 \label{constraint}
 \end{align}
  Only those elements of the density matrix for which this criterion is satisfied survive the limit $\lambda\to 0$. 
  
  Now, notice that, when the constraint (\ref{constraint}) is obeyed, the phases in the third line of equation (\ref{scattering_7}) also cancel exactly. 
  The result is, for evolutions from the initial to the final state for which (\ref{constraint}) holds, the final density matrix is free of infrared divergences.  
  Let us examine some of the consequences of this result.
 
 Since, by unitarity, both the incoming and reduced outgoing density matrices must have unit trace, they both must have nonzero diagonal matrix elements.  
 Moreover,  a question about the evolution of a diagonal element of the density matrix element to another diagonal element of the density matrix  is unaffected by the constraint (\ref{constraint}).  This is due to the fact that,  in such a case, $\{p\}=\{p'\}$ and $\{\tilde p\}=\{\tilde p'\}$ and (\ref{constraint}) is automatically satisfied.  
 This is the question that is usually asked in particle physics: what is the probability that the state $|\{p\}\{k\}\rangle\langle\{p\}\{k\}|$ will evolve to the state $|\{p'\}\{k'\}\rangle\langle\{p'\}\{k'\}|$?  This quantity is finite in the limit $\lambda\to 0$ and it is unconstrained by the condition (\ref{constraint}). We have nothing new to say about it.  
 
 On the other hand, if we ask the questions which probe the off-diagonal elements of the density matrix, the $\lambda\to0$ limit can have drastic consequences.
 If we ask what is the probability of the process 
 \begin{align*}
& \frac{1}{\sqrt{2}}\left(|\{p_1\}\{k_1\}\rangle+|\{p_2\}\{k_2\}\rangle\right)\times~\nonumber\\
&\frac{1}{\sqrt{2}}\left(\langle\{p_1\}\{k_1\} | + \langle\{p_2\}\{k_2\} |\right)\implies |\{p'\}\{k'\}\rangle\langle\{p'\}\{k'\}|\,,
 \end{align*}
we find that,  unless $\{p_1\}=\{p_2\}$, it is
\begin{align*}
\frac{1}{2}~{\rm Prob.~of~} |\{p_1\}\{k_1\}\rangle\langle\{p_1\}\{k_1\}|& \implies |\{p'\}\{k'\}\rangle\langle\{p'\}\{k'\}|\nonumber\\
&+\nonumber\\
\frac{1}{2}~{\rm Prob.~of~}  |\{p_2\}\{k_2\}\rangle\langle\{p_2\}\{k_2\} |& \implies |\{p'\}\{k'\}\rangle\langle\{p'\}\{k'\}|\,.
\end{align*}
There is no interference between the incoming states. 

The probability for evolving to a superposition, on the other hand
 \begin{align*}
 |\{p\}\{k\}\rangle\langle\{p\}\{k\}|&\implies \frac{1}{\sqrt{2}}\left(|\{ p_1'\}\{ k_1'\}\rangle+|\{p_2'\}\{k_2'\}\rangle\right)\nonumber\\
 &\times~\frac{1}{\sqrt{2}}\left( \langle\{p_1'\}\{k_1'\} | + \langle\{p_2'\}\{k_2'\} |\right)
\end{align*}
 unless   $\{p_1'\}=\{p_2'\}$, is simply the sum
 \begin{align*}
\frac{1}{2}~{\rm Prob.~of~} |\{p\}\{k\}\rangle\langle\{p\}\{k\}| & \implies |\{p_1'\}\{k_1'\}\rangle\langle\{p_1'\}\{k_1'\}|\nonumber\\
&+\nonumber\\
\frac{1}{2}~{\rm Prob.~of~}  |\{p\}\{k\}\rangle\langle\{p\}\{k\}| &\implies |\{p_2'\}\{k_2'\}\rangle\langle\{p_2'\}\{k_2'\}|\,.
 \end{align*}
 The state experiences complete decoherence.

 In order to get the results outlined above, we reduced the final state density matrix by tracing it over the Fock space states
 of the soft scalar fields.  Of course, since we are tracing over this entire subspace of the total Hilbert space, 
 tracing over any redefinition of the Fock basis for soft scalars by a unitary transformation 
 must give the same answer.  To find a result that is any different than what we have obtained, 
 one would have to use a redefinition of the basis that is not implemented by a proper unitary transformation. 
 Of course, it is easy to find such improper unitary transformations in a Fock space.  
 Here, the relevant one is the basis constructed around  certain coherent states which become improper coherent states when the
 fundamental infrared cutoff is removed.  Indeed, 
this redefinition of the basis for soft scalar states is easily implemented and, when one subsequently traces in such a basis, the reduced final state density matrix differs in ways that are physically consequential.  In that basis the evolution does not exhibit  the severe  
decoherence or suppression of interference that we found for the Fock state basis. 
 
 However, then we would be in a situation where the $S$ matrix evolves an initial soft scalar Fock  vacuum to soft scalar coherent states which live in a different Hilbert space and the $S$ matrix itself is therefore not a proper unitary operator.  The only way to preserve  unitarity of  $S$ in this context is to also use the coherent states as incoming states, so that $S$ evolves coherent states to coherent states in such a way that it is unitary.    This is the gist of what is done in the dressed state formalism which we will discuss in the context of soft scalar fields in the next section. We expect that the dressed states, in the limit where the cutoff is removed, will have the same problems with violations of Lorentz invariance as photon and graviton dressed states \cite{Froehlich1979,Froehlich1979a,Balachandran2013}.   
\section{ Soft Scalar Dressing} \label{sec: dressing}

  In quantum electrodynamics, a dressed state \cite{Chung1965,Kibble1968} is a modification of the  quantum state of a charged hard particle which attaches a coherent state of soft photons to it.  As well, it must be accompanied by a singular redefinition of the phases of the $S$ matrix \cite{Kulish1971}. A similar idea can be used to obtain dressed states of gravitationally charged particles in quantum gravity when that theory is written as an effective field theory for perturbations of flat spacetime \cite{Ware2013,Choi2019}. In both electrodynamics and gravity, the dressing of states can be done in such a way that the $S$ matrix that describes the scattering of hard dressed particles is infrared finite. In both cases, it has been argued that the soft photons or gravitons simply decouple from the dressed states \cite{Kulish1971,Mirbabayi2016}
  in that their interactions are suppressed by factors of $E_{res}/$(hard particle scales) which can be very small.
  In the following, we will argue that the same dressing procedure can be implemented for hard particles which interact with soft scalar fields. In our simple model (\ref{lagrangian_density}) it is a close parallel to the construction for photons or gravitons.  We will reserve the discussoin of decoupling or non-decoupling for a later section. 
 
Consider an incoming Fock space state of hard particles $ |\{p\}\{k\}\rangle$. We will consider our quantum field theory with a fundamental infrared cutoff $\lambda$. 
 Following Chung and Faddeev and Kulish \cite{Chung1965,Kulish1971}, we define the dressed state, which we denote by $ |\{p\}\{k\}\rangle\rangle$,  as   
 \begin{align}\label{dressing}
  |\{p\}\{k\}\rangle\rangle~\equiv~ W(\{p\}) ~ |\{p\}\{k\}\rangle\,
 \end{align}
 where $W(\{p\})$ is the unitary operation
implemented on an incoming state with fermion quantum numbers $\{p\}$ as
 \begin{align}
& W(\{p\})=\exp\left( R(\{p\} ) \right) \label{dressing1} \\
 &R(\{p\})= \int_\lambda^{E_{res}}\frac{d^3k}{\sqrt{(2\pi)^32|\vec k|}}\ \biggl\{
\left[\sum_{p_n\in\{p\}} f(k,p_n)\right] a^\dagger(k)\nonumber\\
&~~~~~~~~~~~~~~~~~~~~~~~~~~~-\left[ \sum_{p_n\in\{p\}} f^*(k,p_n)\right] a(k)\biggr\}  \label{dressing2} \\
&f(k,p)=\frac{gm}{k\cdot p}\,. \label{dressing3}\end{align}
Note that $R(\{p\})$ is anti-Hermitian and $W(\{p\})$ is unitary. 
  In this expression, $\lambda$ is a fundamental infrared cutoff, and  $E_{res}$ is a second cutoff analogous to the detector resolution 
  of the previous section, and we use the same symbol for it here \footnote{Note that $f$ defined in \eqref{dressing3} is real. Thus $f^*$ in  \eqref{dressing2} appears to over-complicate the expression. However, we prefer to write it this way to allow for any (possibly complex) subleading term in the definition of the dressed states, should we require it.}.  
  
   Our normalization of the creation and annihilation operators is such that
  free field is
  \begin{align}\label{phi}
  \phi_{\rm in}(x)=\int \frac{d^3k}{\sqrt{(2\pi)^32|\vec k|}}\left( e^{ikx} a(k)+e^{-ikx}a^\dagger(k)\right)
  \end{align}
  and $$\left[a(k),a^\dagger(k')\right]=\delta^3(\vec k-\vec k')\,.$$ We have omitted the wave-function, $\frac{1}{\sqrt{(2\pi)^32|\vec k|}}$, for the scalar fields from our expressions for the $S$ matrix.  This means that we must compensate by taking the normalization integral for the states to be the Lorentz invariant measure
  $\int \frac{d^3k}{(2\pi)^32|\vec k|}$ when we finally sum over scalar field states.

 In addition to the dressing of states described in equations (\ref{dressing1})-(\ref{dressing3}),  the $S$ matrix must be modified. As  an operator on Fock space, the modified $S$ matrix is similar to the Dyson $S$ matrix that is computed in Feynman-Dyson-Wick perturbation theory and which we used in the previous sections, the only difference is that it should be multiplied by  some phases which take into account the infinite range of interactions. 
If we  consider the transition between a  dressed state $|\{p\}\{k\}\rangle\rangle$  and a dressed state  $|\{p'\}\{k'\}\rangle\rangle$,    modified $S$ matrix, which we shall  denote by the symbol ${\bf S}$ is defined by
\begin{align}
&  {\bf S}  ( \{p\}\{k\};\{p'\}\{k'\})~=~ \langle\langle \{p'\}\{k'\}| ~{\bf S}~|\{p\}\{k\}\rangle\rangle\\
& ~~\equiv~~ e^{i\Phi(\{p\})}~ \langle\langle \{p'\}\{k'\}|~ { S^\lambda}~|\{p\}\{k\}\rangle\rangle~e^{i\Phi(\{p'\})} \label{defn_bf_S}\,;\\
&\Phi(\{p\}) =-\frac{1}{8\pi}\sum_{\substack{p_n,p_m\in \{p\}\\m\neq n}}\frac{g^2m^2}{ \sqrt{(p_n \cdot p_m)^2-m^4}} \ln\frac{E_{res}}{\lambda}\,. 
\end{align}
As we shall see, the infrared diverging phases serve to cancel the infrared divergent parts of the phases due to infrared divergent loop integrals encountered in the computation of $S^\lambda$. 

\section{Infrared  Finiteness of the Dressed $S$ Matrix}\label{cancellation}

In this section, we shall show that matrix elements of the  $\bf S$ matrix between dressed states are free of infrared divergences.
  The proof is very similar to the analogous one for quantum electrodynamics and for perturbative quantum gravity \cite{Chung1965,Ware2013}. Consider the   Dyson $S$ matrix operator $S^\lambda$ computed in renormalized perturbation theory and the dressed states defined in equations (\ref{dressing})-(\ref{dressing3}). 
 The matrix element of the  $\bf S$ matrix between dressed states is given in equation (\ref{defn_bf_S}) which we recopy here for the reader's convenience:
\begin{align}\label{mathcal_S}
&{\bf S}(\{p\}\{k\};\{p'\}\{k'\})\nonumber\\
&= e^{i\Phi(\{p\})}~ \langle\langle \{p'\}\{k'\}|~ { S^\lambda}~|\{p\}\{k\}\rangle\rangle~e^{i\Phi(\{p'\})}\,. 
\end{align}
The superscript $\lambda$ on $S^\lambda$ indicates that the matrix elements  on the right-hand-side of the above equation are computed while using $\lambda$ as a fundamental 
infrared cutoff.  The dressed states $|\{p\}\{k\} \rangle\rangle$ and the phases  $\Phi(\{p\})$ are also defined with this cutoff. Cancellation of this singular dependence 
on $\lambda$ and finiteness as $\lambda$ is put to zero on the right-hand-side of equation (\ref{mathcal_S})  is the ``infrared finiteness'' that we are seeking in this section.

We can use the Baker-Campbell-Hausdorff formula\footnote{For operators $A$ and $B$ with the properties
$[A,[A,B]]=0$ and  $[B,[A,B]]=0$,
 the Baker-Campbell-Hausdorff formula is 
$$
e^A e^B =e^{\frac{1}{2}[A,B]} e^{A+B} = e^{[A,B]} e^B e^A
$$
} to rewrite the exponential operators in the dressed states defined in equations (\ref{dressing})-(\ref{dressing3}) as

\begin{align}
&|\{p'\}\{k'\}\rangle\rangle\nonumber\\
&=
 \exp\biggl\{   \int_\lambda^{E_{res}}\frac{d^3\ell}{ {(2\pi)^32|\vec \ell|} } \biggl[  -\frac{1}{2}
\biggl|\sum_{p_n'\in\{p'\}} f( \ell,p_n ')\biggr|^2\nonumber\\
&+ \sum_{p_n'\in\{p'\}} 
f(\ell,p_n')a^\dagger(\ell)  \biggr] \biggr\}
 |\{p'\}\{k'\}\rangle\,,
\end{align}
\begin{align}
&\langle\langle\{p\}\{k\}|\nonumber\\
&= 
\langle\{p\}\{k\}|
\exp\biggl\{  \int_\lambda^{E_{res}}\frac{d^3\ell}{ {(2\pi)^32|\vec \ell|} }\nonumber\\
 &\times \biggl[  -\frac{1}{2}\biggl|
 \sum_{p_n\in\{p\}} f( \ell,p_n )\biggr|^2- 
 \sum_{p_n\in\{p\}} f^*( \ell,p_n )a(\ell)  \biggr] \biggr\}\,.
  \end{align}
  
Then, using these states, equation (\ref{mathcal_S}) becomes
  \begin{widetext}
\begin{align} 
&{\bf S} (\{p\}\{k\};\{p'\}\{k'\})= \nonumber \\ &\exp\biggl\{ i\Phi(\{p\})+i\Phi(\{p'\})
 + \int_\lambda^{E_{res}}\frac{d^3\ell}{ {(2\pi)^32|\vec \ell|}} \biggl[ -\frac{1}{2}
\left|\sum_{p_n\in\{p\}}f(\ell,p_n)\right|^2 -\frac{1}{2} \left|\sum_{p_n'\in\{p'\}} f(\ell,p_n')\right|^2\ \biggr]  \bigg\} \times   \nonumber
\\
&
\langle\{p\}\{k\}|\exp\biggl\{  - \int_\lambda^{E_{res}}\frac{d^3\ell}{\sqrt{(2\pi)^32|\vec \ell|}}\sum_{p_n\in \{p\}}
f^*(\ell,p_n)a(\ell) \biggr\}   S^\lambda
\exp\biggl\{  \int_\lambda^{E_{res}}\frac{d^3\ell}{\sqrt{(2\pi)^32|\vec \ell|}}\sum_{p_m'\in\{p'\}}
f(\ell,p_m')a^\dagger(\ell)  \biggr\}|\{p'\}\{k'\}\rangle\,.
\end{align}

A compact form for the LSZ formula for the scalar field part of the Dyson $S$ matrix is obtained using a generating functional 
\begin{align*}
\left. S^\lambda =  ~:e^{ \int dz\phi_{\rm in}(z)(-\partial^2)\frac{\delta}{\delta J(z) } }:~\ldots    \mathcal T \ldots e^{ i\int J\phi} \ldots  
 \ldots  \right|_{J=0}
\end{align*}
where $\phi_{\rm in}(x)$ is the asymptotic free field as in equation (\ref{phi}) and $:\ldots:$  denotes the normal ordering.
With this formula, we can find the action of the dressing operator on $S^\lambda$ as
\begin{align*}
&\langle\{p\}\{k\}| e^{ - \int_\lambda^{E_{res}}\frac{d^3\ell}{\sqrt{(2\pi)^32|\vec \ell|}}\sum_{p_n\in \{p\}}
f^*(\ell,p_n)a(\ell) }
S^\lambda
e^{ \int_\lambda^{E_{res}}\frac{d^3\ell}{\sqrt{(2\pi)^32|\vec \ell|}}\sum_{p_m'\in\{p'\}}
f(\ell,p_m')a^\dagger(\ell) } |\{p'\}\{k'\}\rangle
\\ 
&=
 \exp\biggl\{ - \int_\lambda^{E_{res}}\frac{d^3\ell}{{(2\pi)^32|\vec \ell|}}\sum_{ p_n\in\{p\}}f^*(\ell,p_n) \sum_{p_m'\in\{p'\} } f(\ell,p_m') \biggr\}
 \exp\left(\int dzf(z)(-\partial^2)\frac{\delta}{\delta J(z) } \right) \\ & \times \left.
\langle\{p\}\{k\}|~:e^{\int dz\phi_{\text{in}}(z)(-\partial^2)\frac{\delta}{\delta J(z) } }:~\ldots    \mathcal T \ldots e^{ i\int J\phi} \ldots  
 \ldots ~|\{p'\}\{k'\}\rangle~\right|_{J=0}
\end{align*}
In the above formula, we have used the equations
\begin{align}
& e^{ - \int_\lambda^{E_{res}}\frac{d^3\ell}{\sqrt{(2\pi)^32|\vec \ell|}}\sum_{p_n\in \{p\}}
f^*(\ell,p_n)a(\ell) }a^\dagger(k) =
\left( a^\dagger(k)-\sum_{p_n\in\{p\}}f^*(p_n,k)\right) e^{ - \int_\lambda^{E_{res}}\frac{d^3\ell}{\sqrt{(2\pi)^32|\vec \ell|}}\sum_{p_n\in \{p\}}
f^*(\ell,p_n)a(\ell) }\,,
\label{coherent_identity_1}\\
&a(k) e^{ \int_\lambda^{E_{res}}\frac{d^3\ell}{\sqrt{(2\pi)^32|\vec \ell|}}\sum_{p_m'\in\{p'\}}
f(\ell,p_m')a^\dagger(\ell) }
=
e^{ \int_\lambda^{E_{res}}\frac{d^3\ell}{\sqrt{(2\pi)^32|\vec \ell|}}\sum_{p_m'\in\{p'\}}
f(\ell,p_m')a^\dagger(\ell)}\left( a(k)+\sum_{p_n'\in\{p'\}}f(k,p_n')\right)
\label{coherent_identity_2}\,,
\end{align}
and 
 the result is the classical scalar field,  
$$
f(z)=\int_\lambda^{E_{res}} \frac{d^3\ell }{(2\pi)^32|\vec \ell|}\biggl(\sum_{p_n'\in\{p'\}} f(\ell,p_n')e^{i\ell z} - \sum_{p_m\in\{p\}} f^*(\ell,p_m)e^{-i\ell z}\biggr)
$$
occurring in the exponential functional derivative operator in the second line,
$$
\exp\left(\int dz f(z)(-\partial^2)\frac{\delta}{\delta J(z) }\right)\,.$$ 
This operation inserts  ingoing and outgoing soft scalars with wave-functions $\sim f,-f^*$ into the $S$ matrix element
that it operates on.  We can use the soft scalar theorem (\ref{soft1}) to re-write these as the $S$ matrix element without the soft scalars and with an exponential factor,
$$
\exp\biggl\{ \int_\lambda^{E_{res}} \frac{d^3\ell}{(2\pi)^32|\vec \ell |} \left[ \sum_{p_n\in\{p\}} \frac{gm}{p_n\cdot \ell} - \sum_{p_n'\in\{p'\} } \frac{gm}{p'_n\cdot \ell}\right]
\left[ \sum_{p_m\in\{p\} } f(\ell,p_m)-
 \sum_{p_m'\in\{p'\} } f^*(-\ell,p_m') \right]  \biggr\} 
$$
which then leads us to
\begin{align} 
&{\bf S} (\{p\}\{k\};\{p'\}\{k'\})= { S}^\lambda (\{p\}\{k\};\{p'\}\{k'\})~e^{i\Phi(\{p\})+i\Phi(\{p'\})}  \nonumber \\ &\times 
\exp\biggl\{ \int_\lambda^{E_{res}}\frac{d^3\ell}{ {(2\pi)^32|\vec \ell|}}\biggl[ -\frac{1}{2} \left|\sum_{p_n'\in\{p\}}
f(\ell,p_n)\right|^2    -\frac{1}{2}\left| \sum_{p_n'\in\{p'\}}
f(\ell,p_n')\right|^2-\sum_{p_n\in\{p\}}f^*(\ell,p_n) \sum_{p_m'\in\{p'\}} f(\ell,p_m') \nonumber \\ &
+ \biggl[ \sum_{p_n\in\{p\}} \frac{gm}{p_n\cdot \ell} - \sum_{p_n'\in\{p'\} } \frac{gm}{p'_n\cdot \ell}\biggr]
\biggl[ \sum_{p_m\in\{p\} } f(\ell,p_m)-
 \sum_{p_m'\in\{p'\} } f^*(-\ell,p_m') \biggr] \biggr] \biggr\}
 \end{align}
Then, we can use equations (\ref{infrared_cutoff_s_matrix})  to write the above equation as
\begin{align} 
&{\bf S}^\lambda (\{p\}\{k\};\{p'\}\{k'\})= { S}^\Lambda (\{p\}\{k\};\{p'\}\{k'\})~e^{i\Phi(\{p\})-i\bar \Phi(\{p\})+i \Phi(\{p'\})-i\bar \Phi(\{p'\})}  \nonumber \\ 
&\times 
\exp\biggl\{ \int_\lambda^{E_{res}}\frac{d^3\ell}{ {(2\pi)^32|\vec \ell|}}\biggl[ -\frac{1}{2} \left|\sum_{p_n'\in\{p\}}
f(\ell,p_n)\right|^2    -\frac{1}{2} \left|\sum_{p_n'\in\{p'\}}
f(\ell,p_n')\right|^2 -\sum_{p_n\in\{p\}}f^*(\ell,p_n) \sum_{p_m'\in\{p'\}} f(\ell,p_m') \nonumber \\ &
 + \biggl[ \sum_{p_n\in\{p\}} \frac{gm}{p_n\cdot \ell} - \sum_{p_n'\in\{p'\} } \frac{gm}{p_n'\cdot \ell}\biggr]
\biggl[ \sum_{p_m\in\{p\} } f(\ell,p_m)-
 \sum_{p_m'\in\{p'\} } f^*(-\ell,p_m') \biggr] 
    \nonumber \\ &
+ \int_\lambda^{\Lambda}\frac{d^3\ell}{ {(2\pi)^32|\vec \ell|}}\biggl[ -\frac{1}{2} \left( \sum_{p_n\in\{p\}} \frac{gm}{p_n\cdot \ell } \right)^2  
 -\frac{1}{2}\left( \sum_{p_n'\in\{p'\}}\frac{gm}{p_n'\cdot \ell}    \right)^2
 + \sum_{p_n\in\{p\}}\frac{gm}{p_n\cdot \ell}  \sum_{p_m'\in\{p'\}} \frac{gm}{p_m'\cdot \ell }
\biggr]\biggr\}
 \label{almost_final}\end{align}
\end{widetext}
The first line in equation (\ref{almost_final})  contains the phases that come from the Faddeev-Kulish prescription plus from the internal loop contributions to the infrared cutoff $S^\lambda$. The logarithmic infrared divergences cancel in the combinations in which the phases appear there, leaving behind finite parts which we will display shortly.  The second and third lines in equation (\ref{almost_final}) comes from the normalizations and overlaps of the coherent states. The fourth line in equation (\ref{almost_final}) contains the result of using the soft scalar theorem to take into account the soft scalars coming from the coherent states. The last line in equation (\ref{almost_final})  is the contribution of internal loops encountered in the computation of $S^\lambda$.  It is easy to see that the $\lambda$-dependence of the sum of all of these terms cancels when we put $f(p,\ell)=\frac{gm}{p\cdot\ell}$. The latter exponential factors simplify as
\begin{align*} 
&\exp\biggl\{    
 \int_{E_{res}}^{\Lambda}\frac{d^3\ell}{ {(2\pi)^32|\vec \ell|}}\biggl[ -\frac{1}{2} \left( \sum_{p_n\in\{p\}} \frac{gm}{p_n\cdot \ell } \right)^2 \nonumber\\
 &
 -\frac{1}{2}\left( \sum_{p_n'\in\{p'\}}\frac{gm}{p_n'\cdot \ell}    \right)^2
 + \sum_{p_n\in\{p\}}\frac{gm}{p_n\cdot \ell}  \sum_{p_m'\in\{p'\}} \frac{gm}{p_m'\cdot \ell }
\biggr]\biggr\}
\nonumber \\
&=\left( \frac{E_{\rm res}}{\Lambda}\right)^{A(\{p\}\{p'\})}
\end{align*}
where
\begin{align*}
A(\{p\}\{p'\})& = \frac{1}{8\pi^2}\sum_{nm} \frac{g^2m^2\eta_m\eta_n}{\sqrt{(p_m\cdot p_n)^2-m^4}} 
\biggl[i\biggl(\frac{1+\eta_m\eta_n}{2}\biggr)
\nonumber\\
&+ \ln\frac{1+\sqrt{ 1 - \frac{m^4}{ (p_m\cdot p_n)^2} } }{1-\sqrt{ 1 - \frac{m^4}{ (p_m\cdot p_n)^2} } } \biggr]\,.
 \end{align*}
The final result for the element of the $S$ matrix in dressed states is simply
\begin{align}\label{final_formula_for_dressed_S_matrix}
{\bf S}  (&\{p\}\{k\};\{p'\}\{k'\})\nonumber\\
&=\left( \frac{E_{\rm res}}{\Lambda}\right)^{A(\{p\}\{p'\})}~ { S}^\Lambda (\{p\}\{k\};\{p'\}\{k'\})\,.
\end{align}
The left-hand-side of the above formula (\ref{final_formula_for_dressed_S_matrix}) is the matrix element of the modified $S$ matrix, $\bf S$, computed with dressed states. 
It is the amplitude for the transition from dressed state $|\{p\}\{k\}\rangle\rangle$ to dressed state $|\{p'\}\{k'\}\rangle\rangle$. The right-hand-side simply contains the Dyson $S$ matrix, $S^\lambda$, its matrix elements $ S^\Lambda (\{p\}\{k\};\{p'\}\{k'\})$ computed with undressed Fock space states and with an infrared cutoff $\Lambda$.  It is multiplied the  factor $\left( \frac{E_{\rm res}}{\Lambda}\right)^{A(\{p\}\{p'\})}$ made from the ratio of cutoffs raised to a complex, momentum-dependent exponent.  The right-hand-side does not depend on $\Lambda$ in that the $\Lambda$-dependence of $S^\Lambda$ is compensated by the $\Lambda$ dependence of the  factor.  However, the right-hand-side  does depend on the cutoff, $E_{res}$ which is now a parameter of the theory. 
\section{Non-decoupling of Soft Scalar Emission}\label{sec: non-decoupling}

Having constructed the infrared finite $S$ matrix for dressed states, we are now interested in computing the amplitude of emission of additional soft scalars beyond those in
the dressing. The vanishing of such amplitudes and the factorization of the $S$ matrix into hard and soft sectors for the scattering amplitudes of dressed states is 
an already well-known feature of quantum electrodynamics and perturbative quantum gravity \cite{Kulish1971,Mirbabayi2016,Carney2018,Choi2019}. 
There, the factorization has to do with the fact that one could correct the dressing factors to take into account the next-to-leading contributions to the soft theorem.  Then amplitude for the interaction of suitably dressed states to emit or absorb an additional soft photon or soft graviton is indeed suppressed by powers of the detector resolution cutoff $E_{res}$.  This decoupling also has sound physical reasoning.  Photons or gravitons with wavelengths the size of the solar system should have nothing to do with electrodynamic or gravitational physics at a subatomic scale. 

As we shall see, the case of the massless scalar field is a little different.  We will find that soft scalars do not decouple.   
A scattering event for hard dressed particles can produce soft scalars with an amplitude of the same order as other radiative corrections to the amplitude and at the same order as the interactions between the hard particles themselves.  We will outline the argument for this in the following.   Later, in the next section, we will examine the  entanglement of the dressed hard particles and the soft particles that are produced in the scattering of hard particles. 

Let us consider an outgoing  dressed state which contains an additional soft scalar 
\begin{align}
|\{p'\}\{k'\}q'\rangle\rangle
  =e^{R(\{p'\})} a^\dagger(q') |\{p'\} \{k'\} \rangle  
\end{align}

 We would like to compute the dressed $S$ matrix element ${\bf S}( \{p\}\{k\};\{p'\}\{k'\}q')$ which is defined
 as the quantity
 \begin{align}
 &{\bf S}( \{p\}\{k\};\{p'\}\{k'\}q')=
 e^{i \Phi(\{p\}) }\times \nonumber\\
 & \langle\{p\}\{k\}|W^\dagger (\{p\})S^\lambda W(\{p'\}) a^\dagger(q')|\{p'\}\{k'\}\rangle e^{i \Phi(\{p'\})}
 \end{align}
 We can move the scalar creation operator past the operator $W(\{p'\})$ by using the identity (\ref{coherent_identity_1}) to get 
  \begin{align}
 &{\bf S}( \{p\}\{k\};\{p'\}\{k'\}q')
 \nonumber \\ &
 =e^{i \Phi(\{p\}) + i \Phi(\{p'\}) } \langle\{p\}\{k\}|W^\dagger (\{p\})S^\lambda\nonumber\\
 &\times\left( a^\dagger (q')- \sum_{p_n'\in\{p'\} }f(p'_n,q')\right)W(\{p'\}) )|\{p'\}\{k'\}\rangle
 \end{align}
 Then, the $a^\dagger(q')$ can either act on $W^\dagger(\{p\})$, for which we use equation (\ref{coherent_identity_2})  or it could be absorbed into the $S$ matrix, in which case we can use the soft scalar theorem (\ref{sub_soft_theorem}), which we will keep to next-to-leading order. The result is a cancellation between the singular factors and the $f$'s, leaving only the contributions from the next-to-leading soft scalar theorem,
  \begin{align}
 &{\bf S}( \{p\}\{k\};\{p'\}\{k'\}q')
 = ~\left( \frac{E_{\rm res}}{\Lambda}\right)^{A(\{p\}\{p'\})}\nonumber\\
 &\times  \left\{-\sum_{p_n\in\{p\}\{p'\} }\left(\frac{  g}{2m}\right)- g\hat\partial_m\right\}{ S^\Lambda}( \{p\}\{k\};\{p'\}\{k'\})
 \end{align}
 where quantities on the right-had-side are defined in the discussions around equation (\ref{sub_soft_theorem})-(\ref{sub_soft_theorem_1}) and (\ref{final_formula_for_dressed_S_matrix}).
 
 This equation is one of our central results.  The right-hand-side, unlike what occurs for photons or gravitons, cannot be written in a form that depends only on the initial and final states in a way that its effect can be absorbed into  $W(\{p'\})$ or  $W^\dagger(\{p\})$. Even absorbing the first term with $-\frac{g}{2m}$ for each external fermion would modify $W$ in such a way that it is no longer unitary.   It is easy to confirm by a simple tree level computation that $-g\hat\partial_m$ operating on an amputated correlation function simply cannot in general be written as something that depends only on $\{p\}\{k\}$ plus something that depends only on $\{p'\}\{k'\}$ and its action  therefore cannot be absorbed into the $W$'s either.  Our conclusion is that the soft scalars couple to the dressed $S$ matrix at order $g$ in the coupling constant, which makes them just as coupled as the other particles.
\section{Conclusions}\label{conclusion}
 
 We have shown that, in the context of our admittedly rather specialized model, the infrared problem due to massless scalar fields is practically identical to that for photons in quantum electrodynamics and for gravitons in perturbative quantum gravity.  
 This is in spite of the fact that there are no apparent internal symmetries, either continuous or discrete, whatsoever.   
 This means that there are no conserved Noether currents beyond the energy-momentum tensor, no Ward-Takahashi identities and no apparent asymptotic symmetries.
 We have found what we conjecture is different in the soft scalar theory.  The difference lies in the next-to-leading soft scalar theorem.  Unlike the case of photons or gravitons where Ward-Takahashi identities help to write those terms as referring only to the initial and final states \cite{Choi2019}, their scalar analog cannot be written that way. Then, unlike for photons and gravitons, the next-to-leading behaviour cannot be absorbed by modifying the hard particle dressing.  Soft scalars are still produced when dressed hard particles interact.
 
 Now that we have demonstrated that soft scalars do not decouple from the interactions of dressed states, we can revisit the question as to whether they carry any significant amount of information.  We still expect the scenario where the interactions of hard particles in a scattering event also produces a cloud of soft particles and those soft particles are undetectable.  Our experimental resources only have access to the dressed hard particles. To proceed, we could simply ask the same question that we did for scattering of undressed hard particles.  We consider an incoming density matrix state
\begin{align}  \sum_{\substack{ \{p\}\{k\}  \\ \{\tilde p\} \{\tilde k\}  } }  |\{p\}\{k\}\rangle\rangle \rho_{\{p\}\{k\};\{\tilde p\}\{\tilde k\}}~\langle\langle\{\tilde p\}\{\tilde k\}| \,\nonumber\\
\end{align}
which should evolve to an outgoing density matrix
 \begin{align}
& ~ \sum_{\substack{ \{p\}\{k\}\ \\ \{\tilde p\} \{\tilde k\}  } } ~ \sum_{\substack{ \{p'\}\{k'\}\{q'\}  \\ \{\tilde p'\} \{\tilde k'\} \{\tilde q'\} } }  |\{p'\}\{k'\}\{q'\}\rangle\rangle \langle\langle\{\tilde p'\}\{\tilde k'\}\{\tilde q'\} | 
\nonumber \\ &
 ~~~~~~~ \times  {\bf S}^{ \dagger}(  \{p'\}\{ k'\}\{q\} ; \{p\}\{k\})~ \rho_{\{p\}\{k\};\{\tilde p\}\{\tilde k\}}~\nonumber\\ &
 ~~~~~~~ \times  {\bf S}( \{\tilde p\}\{\tilde k\} ;   \{\tilde p'\}\{\tilde k'\}\{ q'\} )\,
\end{align}
which we now trace over the outgoing soft scalars to obtain a reduced density matrix which describes all of the accessible physics of the hard particles in the outgoing state,
 \begin{align}
&\rho_{\rm final}=  ~ \sum_{\substack{ \{p\}\{k\}\ \\ \{\tilde p\} \{\tilde k\}  } } ~ \sum_{\substack{ \{p'\}\{k'\} \\ \{\tilde p'\} \{\tilde k'\} } }  |\{p'\}\{k'\}\rangle\rangle \langle\langle\{\tilde p'\}\{\tilde k'\} | 
\nonumber \\ &
 ~~~~~~~ \times\sum_{\{q\}}  {\bf S}^{ \dagger}(  \{p'\}\{ k'\}\{q\} ; \{p\}\{k\})~ \rho_{\{p\}\{k\};\{\tilde p\}\{\tilde k\}}~\nonumber \\ &
 ~~~~~~~
  {\bf S}( \{\tilde p\}\{\tilde k\} ;   \{\tilde p'\}\{\tilde k'\}\{ q\} ) \,.
\label{soft_info_decoupling_1}\end{align}
Then, we observe that the trace over the soft scalars involves 
\begin{widetext}
\begin{align}
& \sum_{\{q\}}  {\bf S}^{ \dagger}(  \{p'\}\{ k'\}\{q\} ; \{p\}\{k\})~ \rho_{\{p\}\{k\};\{\tilde p\}\{\tilde k\}}~
  {\bf S}( \{\tilde p\}\{\tilde k\} ;   \{\tilde p'\}\{\tilde k'\}\{ q\} ) \nonumber \\ &
 = {\bf S}^{ \dagger}(  \{p'\}\{ k'\} ; \{p\}\{k\})~ \rho_{\{p\}\{k\};\{\tilde p\}\{\tilde k\}}~
  {\bf S}( \{\tilde p\}\{\tilde k\} ;   \{\tilde p'\}\{\tilde k'\} )
 \nonumber \\
 & +\int_\lambda^{E_{res}}\frac{d^3q'}{(2\pi)^32|\vec q'|}
 {\bf S}^{ \dagger}(  \{p'\}\{ k'\}q'; \{p\}\{k\})~ \rho_{\{p\}\{k\};\{\tilde p\}\{\tilde k\}}~
  {\bf S}( \{\tilde p\}\{\tilde k\} ;   \{\tilde p'\}\{\tilde k'\}q')
  +\ldots
\end{align}
\end{widetext}
Since the combination  
\begin{align*}
&{\bf S}^{ \dagger}(  \{p'\}\{ k'\}q'; \{p\}\{k\}) \rho_{\{p\}\{k\};\{\tilde p\}\{\tilde k\}}
  {\bf S}( \{\tilde p\}\{\tilde k\} ;   \{\tilde p'\}\{\tilde k'\}q') \nonumber\\
  &~\sim~ ({q'})^0
\end{align*}
  is not singular as $q\to 0$, it 
  goes like a constant, the integration in the last line of the above equation produces a factor of $E_{res}^2$ and the higher order terms represented by the ellipses there have multiple volume integrals which produce higher orders of $E_{res}$. All of these are suppressed.  In the approximation where we neglect contributions with positive powers of $E_{res}$,  we neglect such terms.     Then equation (\ref{soft_info_decoupling_1})
  becomes
   \begin{align}
&\rho_{\rm final}=  ~ \sum_{\substack{ \{p\}\{k\}\ \\ \{\tilde p\} \{\tilde k\}  } } ~ \sum_{\substack{ \{p'\}\{k'\} \\ \{\tilde p'\} \{\tilde k'\} } }  |\{p'\}\{k'\}\rangle\rangle \langle\langle\{\tilde p'\}\{\tilde k'\} | 
\nonumber \\ &
 ~~~~~~~ \times  {\bf S}^{ \dagger}(  \{p'\}\{ k'\}  ; \{p\}\{k\})~ \rho_{\{p\}\{k\};\{\tilde p\}\{\tilde k\}}\nonumber \\ &
 ~~~~~~~ \times 
  {\bf S}( \{\tilde p\}\{\tilde k\} ;   \{\tilde p'\}\{\tilde k'\}  ) +\mathcal O(E_{res}^2)
\label{soft_info_decoupling_2} 
\end{align}
with no reference to soft particles at all.  
  This is the sense in which the information contained in the soft particles decouples.  We could produce soft scalars in a scattering experiment.  However, if we do not have the detector resolution to see them directly, we have no way of knowing that they are there. It will also be interesting to understand the implications of our result in the context of  Hawking, Perry, and Strominger's proposal \cite{Hawking2016,Strominger2019} of the resolution of the black hole information paradox which has been heavily criticized based on the decoupling of soft gravitons in the dressed formalism \cite{Mirbabayi2016}. 
  
  We should note that recent work on the memory effect \cite{Prabhu2022} suggests that, as well as the well-known cases of massless QED and Yang-Mills theory, Feddeev-Kulish like dressings may not work for the full nonlinear diffeomorphism invariant quantum gravity. Perhaps studying this issue in the simpler context of trivalently coupled massless scalar fields could shed some light on these complex issues. 
 
\begin{acknowledgments}
This work is supported in part by the Natural Sciences and Engineering Research Council of Canada (NSERC).
\end{acknowledgments}
\bibliography{draft}

\end{document}